# On the feasibility of saltational evolution

Mikhail I. Katsnelson[1,*], Yuri I. Wolf[2], and Eugene V. Koonin[2,*]

[1]Radboud University, Institute for Molecules and Materials, Nijmegen, 6525AJ, Netherlands;
[2]National Center for Biotechnology Information, National Library of Medicine, Bethesda, MD 20894

*For correspondence: m.katsnelson@science.ru.nl; koonin@ncbi.nlm.nih.gov

**Abstract**

Is evolution always gradual or can it make leaps? We examine a mathematical model of an evolutionary process on a fitness landscape and obtain analytic solutions for the probability of multi-mutation leaps, that is, several mutations occurring simultaneously, within a single generation in one genome, and being fixed all together in the evolving population. The results indicate that, for typical, empirically observed combinations of the parameters of the evolutionary process, namely, effective population size, mutation rate, and distribution of selection coefficients of mutations, the probability of a multi-mutation leap is low, and accordingly, the contribution of such leaps is minor at best. However, we show that, taking sign epistasis into account, leaps could become an important factor of evolution in cases of substantially elevated mutation rates, such as stress-induced mutagenesis in microbes. We hypothesize that stress-induced mutagenesis is an evolvable adaptive strategy.

**Significance**

In evolutionary biology, it is generally assumed that evolution occurs in the weak mutation limit, i.e. the frequency of simultaneous fixation of multiple mutations occurring in the same genome and the same generation is negligible. We employ mathematical modeling to show that, although under the typical parameter values of the evolutionary process, the probability of multi-mutational leaps is indeed low, they might become substantially more likely under stress, when the mutation rate is dramatically elevated. We hypothesize that stress-induced mutagenesis in microbes is an evolvable adaptive strategy. Multi-mutational leap might matter also in other cases of substantially increased mutation rate, such as growing tumors or evolution of primordial replicators.

**INTRODUCTION**

'*Natura non facit saltus*' ("nature does not make jumps") is a venerable principle of natural philosophy that was most consistently propounded by Leibniz (1) and later embraced by prominent biologists, in particular, Linnaeus (2). This principle then became one of the key tenets of Darwin's theory that was inherited by the Modern Synthesis of evolutionary biology. In evolutionary biology, the rejection of saltation takes the form of gradualism, that is, the notion that evolution proceeds gradually, via accumulation of "infinitesimally small" heritable changes (3, 4). However, some of the most consequential evolutionary changes, such as, for example, the emergence of major taxa, seem to occur abruptly rather than gradually, prompting hypothesis on the importance of saltational evolution, for example, by Goldschmidt ("hopeful monsters") and Simpson ("quantum evolution"). Subsequently, these ideas have received a more systematic, even if qualitative, treatment in the concepts of punctuated equilibrium (5, 6) and evolutionary transitions (7, 8).

Within the framework of modern evolutionary biology, gradualism corresponds to the weak-mutation limit, that is, an evolutionary regime in which mutations occur one by one, consecutively, such that the first mutation is assessed by selection and either fixed or purged from the population, before the second mutation occurs (9). A radically different, saltational mode of evolution (10, 11) is conceivable under the strong-mutation limit (9) whereby multiple mutation occurring within a single generation and in the same genome potentially could be fixed all together. Under the fitness landscape concept (12, 13), gradual or more abrupt evolutionary processes can be depicted as distinct types of trajectories on fitness landscapes (Figure 1). The typical evolutionary paths on such landscapes are thought to be one step at a time, uphill mutational walks (12). In small populations, where genetic drift becomes an important evolutionary factor, the likelihood of downhill movements becomes non-negligible (14). In principle, however, a different type of moves on fitness landscapes could occur, namely, leaps (or "flights") across valleys when a population can move to a different area in the landscape, for example, to the slope of a different, higher peak, via simultaneous fixation of multiple mutations (Figure 1).

We sought to obtain analytically, within the population genetics framework, the conditions under which multi-mutational leaps might be feasible. The results suggest that, under most typical parameters of the evolutionary process, leaps cannot be fixed. However, taking sign epistasis into account, we show that saltational evolution could become relevant under conditions of elevated mutation rate under stress so that stress-induced mutagenesis could be considered an evolvable adaptation strategy.

**RESULTS**

**Multi-mutation leaps in the equilibrium regime**

Let us assume (binary) genomes of length $L$ (in the context of this analysis, $L$ should be construed as the number of evolutionarily relevant sites, such as codons in protein-coding genes, rather than the total number of sites), the probability of single mutation $\mu \ll 1$ per site per round of replication (generation), and constant effective population size $N_e \gg 1$. Then, the transition probability from sequence $i$ to sequence $j$ is (Ref. (15), Eq.3.11):

$$q_{ij} = \mu^{h_{ij}}(1-\mu)^{L-h_{ij}} \tag{1}$$

where $h_{ij}$ is the Hamming distance (number of different sites between the two sequences). The number of sequences separated by the distance $h$ is equal to the number of ways $h$ sites can be selected from $L$, that is,

$$N_h = \frac{L!}{(L-h)!h!} \approx \frac{L^h}{h!} \tag{2}$$

where the last, approximate expression is valid under the assumption that $L \gg 1$ and $L \gg h$ ($h$ can be of the order of 1).

Assuming also $\mu \ll 1$, we obtain a typical combinatorial probability of leaps over the distance $h$:

$$Q(h) \sim N_h q(h) = P_h(L\mu) \equiv \frac{(L\mu)^h e^{-L\mu}}{h!} \tag{3}$$

which is a Poisson distribution with the expectation $L\mu$.

In steady state, the probability of fixation of the state $i$ is proportional to $\exp(-v x_i)$ where

$$v = N_e - 1, 2(N_e - 1), 2N_e - 1 \tag{4}$$

for the Moran process, haploid Wright-Fisher process, and diploid Wright-Fisher process, respectively, and $x_i = -\ln f_i$ where $f_i$ is the fitness of the genotype $i$ ($x_i$ is analogous to energy in the Boltzmann distribution within the analogy between population genetics and statistical physics (16)). Then, the rate of the occurrence and fixation of the transition $i \to j$ is (15)

$$W_{ij} = q_{ij} \frac{v(x_i - x_j)}{\exp[v(x_i - x_j)] - 1} \tag{5}$$

The distribution function of the fitness differential $\Delta_{ij} = x_i - x_j$ has to be specified (hereafter, we refer to $x$ as fitness, omitting logarithm for brevity). We analyze first the case without epistasis, that is, with additive fitness effects of individual mutations:

$$\Delta(h) = y_1 + y_2 + \cdots + y_h \tag{6}$$

where $y_i$ are independent random variables with the distribution functions $G_j(y_j)$. Then, the distribution function of the fitness difference is

$$\rho_h(\Delta) = \prod_j \int dy_j G_j(y_j) \delta\left(\sum_j y_j - \Delta\right) = \int_{-\infty}^{\infty} \frac{dk}{2\pi} e^{-ik\Delta} \prod_j \int dy_j e^{iky_j} G_j(y_j) \tag{7}$$

which is obtained by using the standard Fourier transformation of the delta-function.

Now, let us specify the distribution of the fitness effects of mutations $G_i(y_i)$, assuming an exponential dependency of the probability of a mutation on its fitness effect, separately for beneficial and deleterious mutations:

$$P_i(y_i) = \begin{cases} D_i e^{-\epsilon_i y_i}, & y_i > 0 \\ D_i r_i e^{\epsilon_i y_i}, & y_i < 0 \end{cases} \tag{8}$$

where $D_i$ is the normalization factor, $r_i$ is the ratio of the probabilities of beneficial and deleterious mutations, and $\epsilon_i$ is the inverse of the characteristic fitness difference for a single mutation (see below). For simplicity, we assume here the same decay rates for the probability density of the fitness effects of beneficial and deleterious mutations. Empirical data on the distributions of fitness effects of mutations (17, 18) clearly indicate that $r_i \ll 1$. From the normalization condition,

$$D_i = \frac{\epsilon_i}{1+r_i} \approx \epsilon_i \tag{9}$$

Note that the mean of the fitness difference (selection coefficient) when the distribution of the fitness effects is given by (8) is

$$|s_i| = \int dy_i G_i(y_i) y_i \approx \frac{1}{\epsilon_i} \tag{10}$$

For simplicity, we start with an assumption that the values of $D_i$ and $r_i$ are independent of $i$. For the model (8):

$$\int_{-\infty}^{\infty} dy\, G(y) e^{iky} = iD\left(\frac{1}{k+i\epsilon} - \frac{r}{k-i\epsilon}\right) \tag{11}$$

Then, from equation (5), the rate of fixation is equal to

$$\varphi(h) = \int_{-\infty}^{\infty} d\Delta \frac{\nu\Delta}{e^{\nu\Delta}-1} \rho_h(\Delta) \tag{12}$$

Substituting (11) into (7), we obtain

$$\rho_h(\Delta) = -\frac{i^{h+1}\epsilon^h}{(h-1)!(1+r)^h} \frac{d^{h-1}}{dk^{h-1}}\left[\left(1 - r\frac{k+i\epsilon}{k-i\epsilon}\right)^h e^{-ik\Delta}\right]_{k=-i\epsilon}, \Delta > 0$$

$$\rho_h(\Delta) = \frac{i^{h+1}\epsilon^h}{(h-1)!(1+r)^h} \frac{d^{h-1}}{dk^{h-1}}\left[\left(\frac{k-i\epsilon}{k+i\epsilon} - r\right)^h e^{-ik\Delta}\right]_{k=i\epsilon}, \Delta < 0 \tag{13}$$

Consider first the case $r = 0$ (all mutations are deleterious). Then, $\rho_h(\Delta < 0) = 0$. For $\Delta > 0$, that is, decrease of the fitness, we have:

$$\rho_h(\Delta) = \frac{\Delta^{h-1}\epsilon^h}{(h-1)!} e^{-\epsilon\Delta} \tag{14}$$

Then, the fixation rate (12) of a leap at a distance $h$ is equal to

$$\varphi(h) = \frac{z^h}{(h-1)!} \int_0^{\infty} dt \frac{t^h e^{-zt}}{e^t - 1} = hz^h \zeta(h+1, z+1) \tag{15}$$

where $z = \epsilon/\nu$ and $\zeta(x, y)$ is the Hurwitz zeta function $\zeta(x, y) = \sum_{k=0}^{\infty} \frac{1}{(k+y)^x}$. Therefore, the rate of fixation for leaps of the length $h$ is equal to $W(h) = P_h(L\mu)\varphi(h)$.

In one extreme, if $z \gg 1$ ($\nu|s| \ll 1$, neutral landscape), $\varphi(h) \approx 1$ and mutations are fixed at the rate they occur. In the opposite extreme case of strong negative selection ($z \ll 1$, $\nu|s| \gg 1$), $\varphi(h) \approx hz^h\zeta(h+1)$ where $\zeta(x)$ is the Riemann zeta function. For a rough estimate, $\zeta(h+1)$ can be replaced by 1, and then, $W(h) \approx L\mu e^{-L\mu} P_{h-1}(L\mu z)$. In this case, the maximum of $W(h)$ is reached at $h = L\mu z \cong \frac{L\mu}{\nu|s|}$ which gives a non-negligible fraction of multi-mutation leaps ($h > 1$) among the fixed mutations only for $L\mu \geq \nu|s|$. However, in this case, the value of $W(h)$ at this maximum is exponentially small because $e^{-L\mu} < e^{-\nu|s|}$. Therefore, in the regime of strong selection against deleterious mutations and at high mutations rates ($L\mu \geq \nu|s|$), multiple mutations actually dominate the mutational landscape, but their fixation rate is extremely low. Qualitatively, this conclusion seems obvious, but we now obtain the quantitative criteria for what constitutes "strong selection". We find that, even for $\nu|s|\sim 10$, the rate of multi-mutation leaps ($h = 4$) can be non-negligible ($>10^{-4}$ per generation, Figure 2A) at the optimal $L\mu$ values, whereas for $\nu|s|\sim 100$, any leaps with $h>1$ are unfeasible (Figure 2B).

Under a more realistic model, all values of $\epsilon_i$ (fitness effects of mutations) are different. For $\Delta > 0$ and $r = 0$ (no beneficial mutations)

$$\rho_h(\Delta) = \int_{-\infty}^{\infty} \frac{dk}{2\pi} e^{-ik\Delta} \prod_j \frac{i\epsilon_j}{i\epsilon_j + k} \tag{16}$$

For example, in Kimura's neutral evolution model (19), $\epsilon_i$ is a binary random variable that takes a value of $\infty$ ($|s_i| = 0$, neutral mutation), with the probability $f$, and a value of 0 ($|s_i| = \infty$, lethal mutation), with the probability $1 - f$. Then, $\rho_h(\Delta) = f^h \delta(\Delta)$, $\varphi(h) = f^h$ and $L\mu$ is replaced with $Lf\mu$ in equation (3), a trivial replacement of the total genome length $L$ with the length of the part of the genome where mutations are allowed, $Lf$. Accordingly, $W(h) = P_h(L\mu f)$, and multi-mutation leaps become relevant for $L\mu f \geq 1$.

Let us now estimate the probability of leaps with beneficial mutations ($\Delta < 0$). Assuming $r \ll 1$ (rare beneficial mutations), equation (13) takes the form

$$\rho_h(\Delta) \approx \frac{hr\epsilon}{2^{h-1}} e^{-\epsilon|\Delta|} \tag{17}$$

and the fixation rate of beneficial mutations is

$$\varphi(h) = \frac{hr}{2^{h-1}} z\zeta(2, z) \tag{18}$$

If $z \gg 1$ (weak positive selection), $z\zeta(2, z) \approx 1$, so that the role of beneficial mutations is negligible. If $z \ll 1$ (strong positive selection),

$$\varphi(h) = \frac{hr}{2^{h-1}} \frac{1}{z} \tag{19}$$

Comparing equation (19) with the result for $\Delta > 0$ (equation (18)), one can see that, in this case, beneficial mutations are predominant among the fixed mutations if

$$r > (2\epsilon/v)^{h+1} \tag{20}$$

In this regime, multi-mutation leaps ($h > 4$), occur at non-negligible rates under sufficiently high (but not excessive) mutation rates (Figure 3).

The model considered above assumes independent effect of different mutations (no epistasis, "ideal gas of mutations" model). Now, let us take into account epistasis. In the case of strong epistasis, effects of combinations of different mutations are increasingly strong, diverse and, effectively, unpredictable, resulting in a ragged fitness landscape (20). In the limit of epistasis strength and unpredictability, epistasis creates numerous highly beneficial combinations that, once they occur, are highly likely to be fixed, and a far greater number of highly deleterious combinations that are immediately lethal. Due to the effective randomness of genetic interactions, we consider the resulting landscape as essentially random for $h > 1$, with the frequency of the beneficial combinations $r$ independent of $h$. In this case, the rate of fixation of leaps of the length $h > 1$ is simply

$$W(h) = Q(h)f \sim P_h(L\mu)r \tag{21}$$

If all single mutations ($h = 1$) are deleterious ($v|s| \gg 1$), their rate of fixation (equation (15)) can be approximated by $W(1) = P_1(L\mu) \frac{1}{v|s|}$, whereas for all leaps of the length $h > 1$, the fixation rate is $W(h > 1) = (1 - P_0(L\mu) - P_1(L\mu))f$. Therefore, the condition for $W(h > 1) > W(1)$ is

$$r > \frac{L\mu e^{-L\mu}}{(1-(L\mu+1)e^{-L\mu})} \frac{1}{v|s|} \qquad (22)$$

In the high mutation regime ($L\mu \gg 1$), multiple mutations occur many orders of magnitude more frequently than single mutations, overwhelming the difference of scale between $r$ and $\frac{1}{v|s|}$, and making multi-mutation leaps much more likely. Around the Eigen threshold ($L\mu \approx 1$) (21), the condition corresponds to $r > \frac{1}{(e-2)} \frac{1}{v|s|}$, i.e. the frequency of beneficial multi-mutation combinations should be greater than the reciprocal of the strength of negative selection against individual mutations. In the low mutation regime ($L\mu \ll 1$), the balance between single and multiple mutations tends to $\frac{1}{L\mu} - 1$ and the condition for the dominance of multi-mutation leaps becomes $r > (\frac{1}{L\mu} - 1) \frac{1}{v|s|}$, i.e. the frequency of beneficial multi-mutation combinations should additionally compensate for the excess of single-mutation events.

**Non-equilibrium model of stress-induced mutagenesis**

The analysis presented above suggests that the necessary condition for fixation of multi-mutational leaps is the high mutation regime. At low mutation rates ($L\mu \ll 1$), multi-mutation ($h > 1$) events occur too rarely to be fixed in realistic settings even if the frequency of beneficial combinations among them is reasonably high. However, in the high mutation regime ($L\mu \gg 1$), the above analysis is problematic for two reasons. First, the expression for the fixation rate (equation (5)) is technically valid only for the case when the new mutation is either fixed or lost before the emergence of the next one, which implies $L\mu < 1/v \ll 1$. Second, under any realistic model of the fitness landscape, most mutations should be deleterious. Thus, $L\mu > 1$ implies that most of the progeny carries one or more mutations, and therefore, suffers from these deleterious effects. Under these conditions, the assumption of constant $N_e$ is unrealistic, because the size of such a population will decrease under the mutational load, down to an eventual crash.

The complete analysis of the behavior of a variable-size population under the high mutation regime and strong mutational effects is currently beyond the state of the art. Therefore, here we analyze a simplified model of the short-term behavior of a (microbial) population after the onset of stress-induced mutagenesis ($L\mu \gg 1$).

Consider a microbial population consisting of $N_0$ individuals. Under normal conditions, the population is at an equilibrium, so that approximately $N_0/2$ individuals survive the average generation span and produce $N_1 \approx N_0$ progeny by division (here we consider simple asexual division as the progeny-generating process; other demographic models can be accommodated without loss of generality). The normal mutation rate is low ($L\mu_0 \approx 1/N_e \approx 1/N_0$, according to (22, 23)), so the population can be considered homogeneous. Upon the onset of unfavorable conditions, the survival rate of the wild type individuals drops to $f_w \ll 1/2$ and the mutation rate in the stressed individuals increases such that to $L\mu > 1$.

If $f_w$ is not too small ($f_w N_0 \gg 1$), the immediate wild-type survivors produce $2f_w N_0$ first-generation progeny. With the expected number of mutations per descendant being $L\mu$, the distribution of the number of mutations in the progeny is given by the Poisson distribution with the expected number of mutants with $h$ mutations of $2f_w P_h(L\mu)N_0$.

Let us consider a mutation landscape that is dominated by deleterious mutations with strong sign epistasis. All single mutations are deleterious, so the survival of their carriers over the generation time is $f_1 \ll f_w$. An overwhelming majority of multi-mutation combinations have even stronger negative effects, so for $h > 1$, $f_h \ll f_1$. Some small fraction $r_h$ of these combinations, however, is strongly beneficial in the new conditions, conferring to their carriers the survival rate of ~1/2.

What should the $r_h(h)$ function look like? Intuitively, $r_h(h)$ should decay to 0 at large $h$, or at least, not grow, as it is overwhelmingly likely that a sufficiently large set of mutations would contain a subset that it unconditionally lethal. Here, for simplicity, we consider a general form of $r_h(h)$ that is equal to 0 for $h = 1$ and monotonically decays with $h$ from $r_2$ at an arbitrary rate.

If the deleterious effect of mutations is strong enough ($0 \approx f_h \approx f_1 \ll f_w$), then, the only plausible source of beneficial mutants is the population of wild type individuals (neither single mutants nor multiple mutants that do not carry the beneficial combinations survive to the next generation). The population of the wild type individuals decays exponentially through both the diminished survival and through mutations, reaching $f_w(2f_w P_0(L\mu))^{k-1} N_0$ at the $k$-th generation after the onset of the unfavorable conditions. Ignoring stochastic fluctuations, the total number of wild type individuals that survive until the population collapse can be estimated as

$$N_w^\infty \approx f_w N_0/(1 - 2f_w P_0(L\mu)) \tag{23}$$

which is approximately equal to $f_w N_0$ if $f_w P_0(L\mu) \ll 1/2$.

Over the combined lifetimes of the surviving wild type individuals, the expected number of beneficial mutants

$$E(N_B^\infty) \approx 2N_w^\infty \sum_{h=2}^\infty (r_h P_h(L\mu)) = 2N_w^\infty \sum_{h=2}^\infty \left(r_h \frac{e^{-L\mu}(L\mu)^h}{h!}\right) \tag{24}$$

which depends on the genome-wide mutation rate $L\mu$ and the shape of the $r_h(h)$ function.

Let us first consider the two extreme cases of $r_h(h)$. In the limit of a completely flat function ($r_h = r_2$ for all $h > 2$), equation (23) gives $E(N_B^\infty) \approx 2N_w^\infty r_2(1 - P_0(L\mu) - P_1(L\mu))$. This function asymptotically reaches the value of $2N_w^\infty r_2$ with $L\mu \to \infty$. In the other extreme of a rapidly decaying $r_h(h)$, i.e. $r_h = 0$ for $h > 2$, equation (23) gives $E(N_B^\infty) \approx 2N_w^\infty r_2 P_2(L\mu)$. This function reaches its maximum at $L\mu = 2$ with $E(N_B^\infty) \approx 4N_w^\infty r_2 e^{-2}$.

It can be shown that the estimates for all other monotonically decaying $r_h(h)$ functions reach their maximum at finite values of $L\mu$ with

$$4N_w^\infty r_2 e^{-2} \leq E(N_B^\infty) \leq 2N_w^\infty r_2 \tag{25}$$

Indeed, let us consider first the simplest model $r_h(h) = r_2 \xi^{h-2}$ ($0 < \xi < 1$). Then, equation (24) takes the form

$$E(N_B^\infty) \approx 2N_w^\infty r_2 \xi^{-2} e^{-L\mu}(e^{L\mu\xi} - L\mu\xi - 1) \tag{26}$$

As a function of $L\mu$, the quantity

$$\varphi(L\mu) = e^{-L\mu}(e^{L\mu\xi} - L\mu\xi - 1) \tag{27}$$

reaches the maximum at $L\mu^* = x/\xi$ where $x$ is the solution of the equation

$$\frac{e^x - 1}{x} = \frac{1}{1-\xi} \tag{28}$$

with the value at the maximum

$$\varphi(L\mu^*) = (1-\xi)^{\frac{1-\xi}{\xi}} x\xi(1-\xi+x)^{-1/\xi} \tag{29}$$

For $\xi \ll 1$ (rapid decay of $r_h$), $L\mu^* = 2$ and $E(N_B^\infty) \approx 4N_w^\infty r_2 e^{-2}$. In the opposite limit of slowly decaying $r_h$, $\xi \to 1$, $L\mu^* \approx \ln\frac{1}{1-\xi}$ and $\varphi(L\mu^*) \approx 1$.

For a general slowly decaying function $r_h(h)$, one can find that

$$L\mu^* \approx \ln \left| \frac{r_2}{\left|\frac{dr_2}{dh}\right|_{h \approx L\mu^*}} \right| \tag{30}$$

Importantly, even in this case, the optimal mutation rate $L\mu^*$ increases only logarithmically with the decay rate; furthermore, the optimum value is notably robust to changes in $r_h(h)$ (Figure 4).

The approximate condition for population survival, $E(N_B^\infty) > 2$, can be derived from equations (23) and (25), and is bounded from below by

$$r_2 > \frac{e^2}{2N_0 f_w} \tag{31}$$

at the optimal value of $L\mu$.

**DISCUSSION**

Here, we obtained analytic expressions for the probability of multi-mutation leaps for deleterious and beneficial mutations depending on the parameters of the evolutionary process, namely, effective genome size ($L$), mutation rate ($\mu$), effective population size ($\nu$), and distribution of selection coefficients of mutations ($s$). Leaps in random fitness landscapes in the context of punctuated equilibrium have been previously considered for infinite (24, 25) or finite (26) populations. However, unlike the present work, these studies have focused on the analysis of the dynamics of the leaps rather than on the equilibrium distribution of their lengths. We further address the plausibility of beneficial multi-mutation leaps in the presence of epistasis and outside of equilibrium, e.g. in a microbial population under stress.

The principal outcome of the present analysis are the conditions under which multi-mutation leaps occur at a non-negligible rate in different evolutionary regimes. If the landscape is completely flat (strict neutrality, $s = 0$), the leap length is distributed around $L\mu$, that is, simply, the expected number of mutations per genome per generation. If $L\mu \ll 1$, leaps are effectively impossible, and evolution can proceed only step by step (12). A considerable body of data exists on the values of each of the relevant parameters that define the probability of leaps. Generally, in the long term, the total expected number of mutations per genome per generation has to be of the order of 1 or lower (Eigen threshold) because, if $L\mu \gg 1$, the population ultimately succumbs to mutational meltdown (15, 21, 27). The selection for lower mutation rates is thought to be limited by the drift barrier and, accordingly, the genomic mutation rate appears to be inversely proportional to the effective population size, that is, $L\mu \sim 1/v$ (22, 23). Thus, $L\mu v = const$, which appears to be an important universal in evolution.

To estimate the leap probability under specific values of the relevant parameters, we can use equation (15) and the characteristic values of the parameters, for example, those for human populations. As a crude approximation, $L\mu = 1$, $v = 10^4$, $|s| = 10^{-2}$ which, in the absence of beneficial mutations, translates into the probability of a multi-mutation leap of about $4 \times 10^{-5}$. Thus, such a leap would, on average, require over 23,000 generations which is not a relevant value for the evolution of mammals (given that ~140 single mutations are expected to be fixed during that time as calculated using the same formula). However, short leaps including beneficial mutations can occur with reasonable rates, such as $5 \times 10^{-4}$ for $h = 3$, and the frequency of beneficial mutations $r = 10^{-4}$, so such leaps are only 8 times less frequent than single mutation fixations. Conceivably, such leaps of beneficial mutations could be a minor but non-negligible evolutionary factor. For organisms with $L\mu < 1$ and larger $v$, the probability of leaps is substantially lower than the above estimates, so that under "normal" evolutionary regimes (at equilibrium), the contribution of leaps is negligible.

However, in some biologically relevant and common situations, such as stress-induced mutagenesis, which occurs in microbes in response to double-stranded DNA breaks, the effective mutation rate can locally and temporarily increase by orders of magnitude (28, 29) while the population is going through a severe bottleneck. If the fraction of beneficial combinations of mutations satisfies the condition (31), even in the extreme case when the rest of the mutations are lethal, the population has a chance to survive when its mutation rate ($L\mu$) assumes a value close to the optimum given by equation (30). This value depends on the rate of the decay of the fraction of beneficial combinations of mutations with the number of mutations. Specifically, the optimal value of $L\mu$ equals 2 for the steepest decay of $r(h)$ and increases logarithmically slowly for more shallow functions. Under an extremely severe stress ($N_0 = 10^9$, $f_w = 10^{-3}$), the survival threshold [$r(h)$] corresponds to the fraction of beneficial pairs of mutations of about $3 \times 10^{-6}$. This means that, in the case of a typical bacterial genome of $3 \times 10^6$ base pairs, for each (deleterious) mutation, there is, on average, one other mutation that yields a beneficial combination. This estimate pertains to the extreme case when all individual mutations are highly deleterious. Under more realistic conditions, when many mutations are effectively neutral, and a small fraction is beneficial, the threshold fraction of beneficial combinations will be considerably lower. These estimates indicate that multi-mutation leaps are likely to be an important factor of adaptive evolution under stress. An implication of these findings is that stress-induced mutagenesis could be a selectable adaptive mechanism, however controversial an issue the evolution of evolvability might be (30-34). It should be further noted that, in this situation, large populations will have a higher innovation potential than small ones because the former produce a greater diversity of multi-mutation combinations. In other terms, large populations have a greater chance to cross the entropy barrier to higher fitness genotypes {van Nimwegen, 2000 #1573}. Thus, the stress-induced innovation regime is an alternative to innovation by drift that occurs in small populations (during population bottlenecks) (14, 23).

A different context in which multi-mutation leaps potentially might play a role is evolution of cancers. In most tumor types, mutation rate is dramatically, orders of magnitude elevated compared to normal tissues (35, 36). The effective population size in tumors is difficult to estimate, and therefore, there is not enough information to use the condition (31) to assess the plausibility of multi-mutation leaps. Nevertheless, given the extremely high values of $L\mu$, it cannot be ruled out that the frequency of leaps is non-negligible. Most of the mutations in tumors are passengers that have no effect on cancer progression or exert a deleterious effect (37, 38).

Traditionally, tumorigenesis is thought to depend on several driver mutations that occur consecutively (39, 40). This is indeed likely to be the case in many tumors because the age of onset strongly and positively correlates with the number of drivers (41, 42). However, for a substantial fraction of tumors, no drivers are readily identifiable suggestive of the possibility that, in these cases, tumor progression is driven by 'epistatic drivers' (41), that is, combinations of mutations that might occur by leaps.

Another, completely different area where multi-mutation leaps could be important could be evolution of primordial replicators, in particular, those in the hypothetical RNA World, that are thought to have had an extremely low replication fidelity, barely above the mutational meltdown threshold (21, 43, 44). Furthermore, because the primordial replicators are likely to have been incompletely optimized, the fraction of beneficial mutational combinations could be relatively high. Under these conditions, multi-mutational leaps could have been an important route of evolutionary acceleration and thus might have contributed substantially to the most challenging evolutionary transition of all, that from pre-cellular to cellular life forms.

Taken together, all these biological considerations suggest that multi-mutation leaps with a beneficial effect, the probability of which we show to be non-negligible under conditions elevated mutagenesis, could be an important mechanism of evolution that so far has been largely overlooked. Given that elevated mutation rate caused by stress is pervasive in nature, saltational evolution, after all, might substantially contribute to the history of life, in direct defiance of '*Natura non facit saltus*'.

**Acknowledgements**
MIK acknowledges financial support from the NWO via the Spinoza Prize. YIW and EVK are funded through the Intramural Research Program of the National Institutes of Health of the USA.

**Figures**

**Figure 1. Walks and leaps on different types of fitness landscapes**

Dots show genome states; blue (shirt straight) arrows indicate consecutive moves via fixation of single mutations; red (long curved) arrows indicate multi-mutation leaps.

A. Nearly neutral landscape.

B. Landscape dominated by slightly deleterious mutations.

C. Kimura's model landscape (a fraction of mutations is neutral; the rest are lethal).

D. Landscape combining beneficial and deleterious mutations.

**Figure 2. Rates of leaps on a landscape dominated by deleterious mutations**

Rates of transitions are plotted against the per-genome mutation rate ($L\mu$) and the leap length for different strengths of selection (A: $\nu|s| = 10$ and B: $\nu|s| = 100$). Contour lines indicates orders of magnitude and start from the rate of $10^{-5}$ leaps per generation.

**Figure 3. Rates of leaps on a landscape combining beneficial and deleterious mutations**

Rates of leaps are plotted against the per-genome mutation rate ($L\mu$) and the leap length for different strengths of selection (A and C: $\nu|s| = 10$; B and D: $\nu|s| = 100$) and for different frequencies of beneficial mutations (A and B: $r = 10^{-4}$; C and D: $r = 10^{-3}$). Contour lines indicates orders of magnitude and start from the rate of $10^{-5}$ leaps per generation.

**Figure 4. Abundance of beneficial multi-mutation combinations depending on the mutation rate.**

Abundance of beneficial multi-mutation combinations, $\varphi(L\mu)$, given by equation (24), relative to $r_2$. A: $r_h(h) = r_2(h-1)^{-\alpha}$ with $\alpha = 0$ (blue), $\alpha = 0.5$ (orange), $\alpha = 1$ (green) and $\alpha = 2$ (red). B: $r_h(h) = r_2 \xi^{h-2}$ de with $\xi = 1$ (blue), $\xi = 0.9$ (orange), $\xi = 0.5$ (green) and $\xi = 0.25$ (red).

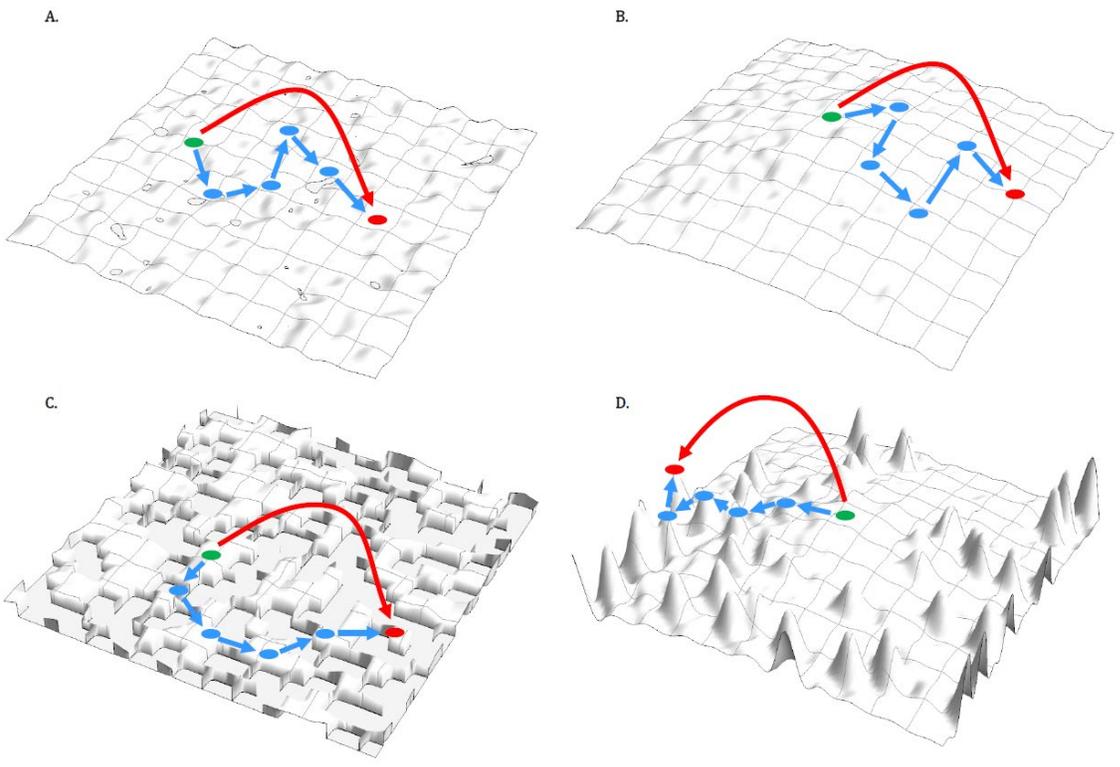

**Figure 1**

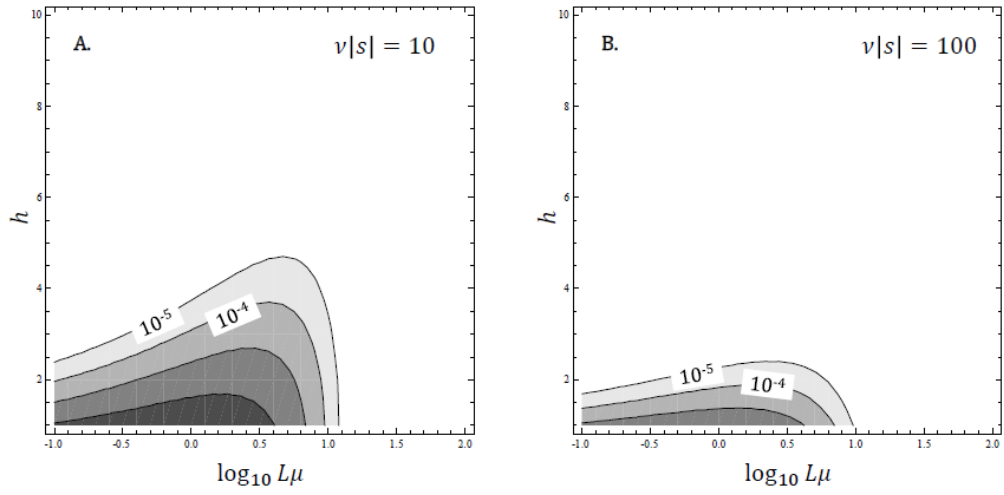

**Figure 2**

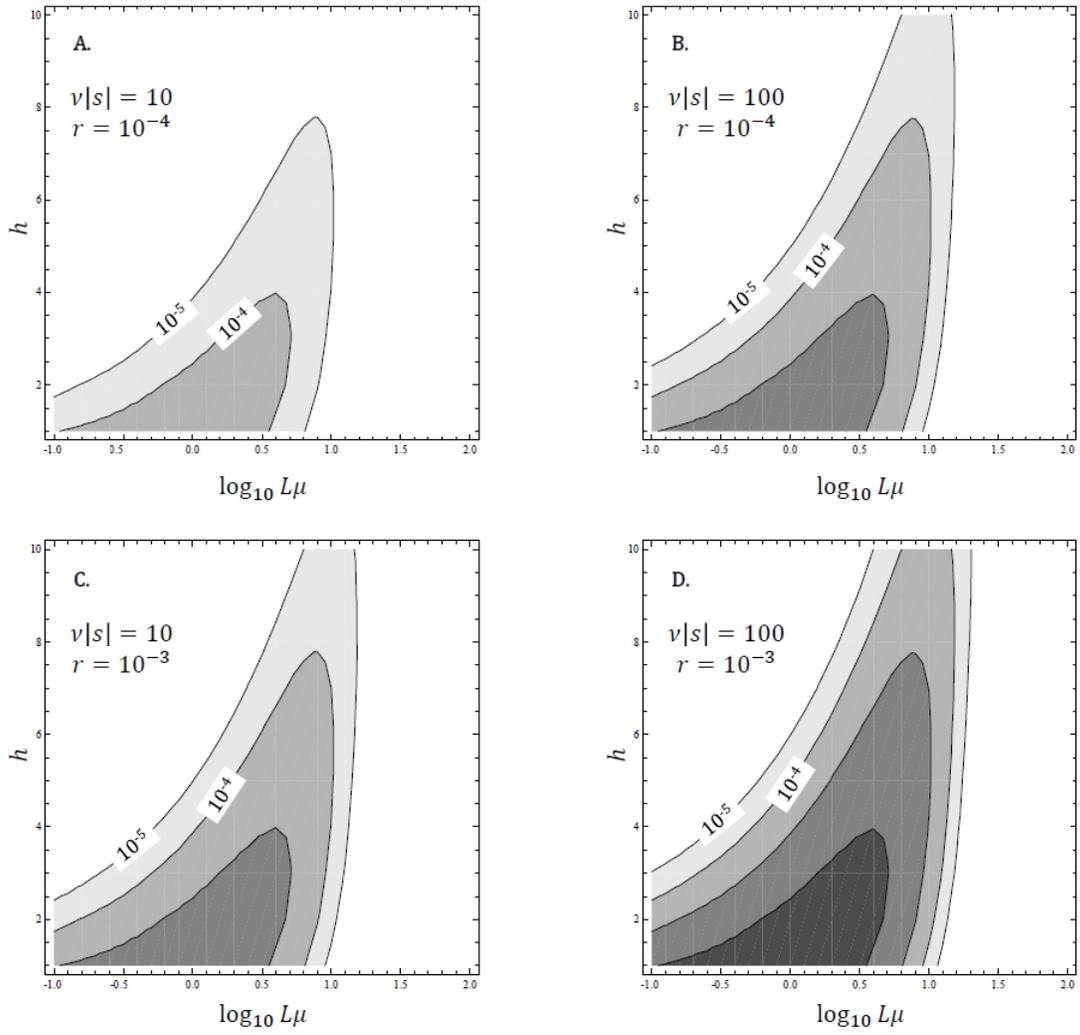

**Figure 3**

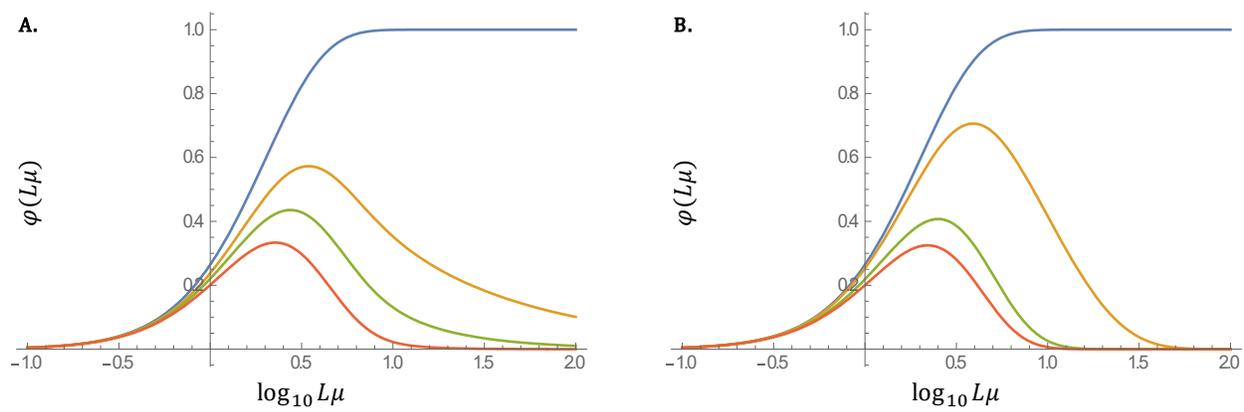

**Figure 4**